\documentclass[preprint,aps,draft,showpacs]{revtex4}
\usepackage{dcolumn}

\begin{document}

\title{Microscopic determination of the nuclear incompressibility\\
within the non-relativistic framework}

\author{G. Col\`o}
\email{colo@mi.infn.it}
\affiliation{Dipartimento di Fisica, Universit\`a degli Studi,
             and INFN sez. di Milano, via Celoria 16, 20133
             Milano (Italy)}
\author{N. Van Giai}
\email{nguyen@ipno.in2p3.fr}
\affiliation{Institut de Physique Nucl\'eaire d'Orsay, IN2P3-CNRS,
             91406 Orsay Cedex (France)}
\author{J. Meyer and K. Bennaceur}
\email{j.meyer@ipnl.in2p3.fr,k.bennaceur@ipnl.in2p3.fr}
\affiliation{Institut de Physique Nucl\'eaire de Lyon, IN2P3-CNRS /
             Universit\'e Claude Bernard Lyon I,
             43, Bd. du 11.11.18, 69622 Villeurbanne Cedex (France)}
\author{P. Bonche}
\email{paul@spht.saclay.cea.fr}
\affiliation{Service de Physique Th\'eorique, CEA Saclay,
             91191 Gif-sur-Yvette Cedex (France)}

\date{\today}

\begin{abstract}

The nuclear incompressibility $K_\infty$ is deduced from
measurements of the Isoscalar Giant Monopole Resonance (ISGMR) in
medium-heavy nuclei, and the resulting value turns out to be model
dependent. Since the considered nuclei have neutron excess, it has
been suggested that the model dependence is due to the different
behaviour of the symmetry energy in different models. To clarify
this issue, we make a systematic and careful analysis based on new
Skyrme forces which span a wide range of values for $K_\infty$,
for the value of the symmetry energy at saturation and for its
density dependence. By calculating, in a fully self-consistent
fashion, the ISGMR centroid energy in $^{208}$Pb we reach, for the
first time within the non-relativistic framework, three important
conclusions: (i) the monopole energy, and consequently the deduced
value of $K_\infty$, depend on a well defined parameter related to
the shape of the symmetry energy curve and called $K_{sym}$; (ii)
Skyrme forces of the type of SLy4 predict $K_\infty$ around 230
MeV, in agreement with the Gogny force (previous estimates using
Skyrme interactions having been plagued by lack of full
self-consistency); (iii) it is possible to build forces which
predict $K_\infty$ around 250 MeV, although part of this increase
is due to our poor knowledge of the density dependence and
effective mass.

\end{abstract}


\maketitle

\section{INTRODUCTION}
\label{intro}

The question about the proper value of the nuclear
incompressibility $K_\infty$ is still open. The model dependence
of this quantity amounts to a difference of the order of $\sim$
10-20\% among the values obtained within different theoretical
models. There is a renewed interest in this issue, 
motivated both by the improved quality of the recent experimental
measurements of the Isoscalar Giant Monopole Resonance (ISGMR),
and by the progress of Relativistic Mean Field models
(RMF) which are to be confronted with more traditional
non-relativistic models based on Skyrme or Gogny effective forces.

Skyrme energy functionals have been widely used in nuclear
structure calculations over three decades. The first Skyrme
effective forces were built in the pioneering work of Vautherin
and Brink\ \cite{Vautherin71}, by fitting their parameters to
nuclear matter properties (the saturation point) and to selected
observables (binding energies and charge radii) of closed-shell
nuclei calculated in the Hartree-Fock (HF) approximation. Later,
many improvements of the Skyrme energy functional have been
devised. These have been possible also because the
mean field approach was extended to the time-dependent case
(time-dependent HF, TDHF) and to its small amplitude limit (Random
Phase Approximation, RPA). Within this scheme, it is possible to
calculate the collective nuclear excitations and to explore the
correlations between their properties and the force parameters, or
some physically meaningful combinations of them. The
relation between the ISGMR and the nuclear incompressibility is
one of such relations.

The introduction of reliable RMF effective Lagrangians is more recent. However,
the progress in this field has been quite fast~\cite{Ring-Bender},
and we can nowadays discuss the properties of the RMF parametrizations 
on the same footing as the Skyrme~\cite{Skyrme} 
and Gogny~\cite{Gogny} functionals.

The nuclear incompressibility $K_\infty$ is related to the
curvature of the energy per particle $E/A$ in symmetric nuclear
matter around the minimum $\varrho_0$, i.e., at the saturation
point:
\begin{equation}
K_\infty \equiv  9\varrho_0^2 {d^2\over d\varrho^2} {E\over A}\vert_{\varrho_0}.
\label{defKinf}
\end{equation}
The interest of determining the value of $K_\infty$ stems also
from its impact on the physics of supernovae and neutron stars.

Until a year ago, the status of the nuclear incompressibility
problem could be summarized as follows. From calculations based on
Skyrme functionals different groups have extracted values of
$K_\infty$ of the order of 210-220 MeV. Using the Gogny
functionals, a value of $K_\infty$ around 230 MeV was obtained.
Finally, the relativistic calculations predicted values in the
range 250-270 MeV. All these results made use of the measured
value of the ISGMR, e.g. in $^{208}$Pb, as explained below.

This situation for the nuclear incompressibility ``puzzle'' has
been reviewed in Ref.\ \cite{Colo03}. There, it was
shown that the
accuracy of the ISGMR data obtained at Texas A\&M\
\cite{Youngblood99} allows for extracting $K_\infty$ with an
experimental error of no more than $\pm$ 12 MeV. Moreover, a rather
important new conclusion was reached. The previous works based on
the Skyrme forces consisted of not fully self-consistent HF plus
RPA calculations in which the two-body residual Coulomb and
spin-orbit forces were neglected. These terms are rather small since 
they affect the monopole energy in $^{208}$Pb by only 4-5\%, but
this produces a change of 8-10\% in the extracted value of the nuclear
incompressibility. By considering this effect, the new value of
$K_\infty$ from Skyrme functionals turns out to be 235 MeV, in very good
agreement with that extracted from the Gogny calculations.
Consequently, there is no discrepancy between the results of the
different non-relativistic calculations. On the other hand, the
gap with the relativistic results remains significant.

The most recent attempts in the literature\
\cite{Piekarewicz02,Vretenar03,Shlomo03,Piekarewicz_tbp} to attack
this problem
are focused on the possible
relation(s) between the monopole energy in systems with a neutron
excess, like $^{208}$Pb, and the density dependence of the
symmetry energy $S(\varrho)$. In fact, one of the clear
differences between the Skyrme and RMF functionals concerns the
behaviour of the symmetry energy around the saturation point
$\varrho_0$. The Skyrme energy functionals are characterized by
smaller values of the symmetry energy at saturation, and of the
corresponding slope, as compared with the RMF functionals. In this
sense it may be said that the RMF functionals are ``stiff''
compared with the ``soft'' non-relativistic ones.

In Ref.~\cite{Piekarewicz02} some effective Lagrangians whose
symmetry energy has different density dependences are built. This
is easy to achieve since, by adjusting the $\rho$-meson coupling
constant, one can at the same time soften the symmetry energy 
$S(\varrho)$ and lower its value at the saturation point, $J
\equiv S(\varrho_0)$. It is thus found that the extracted values
of $K_\infty$ indeed differ and can even become close to the
Skyrme force values if $J$ is around 28 MeV. However, in
Ref.~\cite{Piekarewicz02} no systematic treatment of finite nuclei
is attempted. In Ref.~\cite{Vretenar03} it is pointed out that RMF
parametrizations with $J$ lower than 32 MeV cannot describe
satisfactorily the $N \ne Z$ nuclei. The authors conclude from
their calculations that the lower limit for the RMF value of
$K_\infty$ is around 250 MeV. In Ref.~\cite{Piekarewicz_tbp},
using a markedly improved version of the model
of Ref.~\cite{Piekarewicz02}, this lower limit is confirmed since
$K_\infty$ results to be 248 MeV.

While in the relativistic framework it seems impossible to push
the value of $K_\infty$ below this lower limit, the recent results
of Ref.~\cite{Shlomo03} suggest that one can build at least one
Skyrme-type interaction having $K_\infty$=255 MeV and reproducing
the correct ISGMR energy in $^{208}$Pb. This is at variance with
all other non-relativistic calculations quoted in Ref.~\cite{Colo03}.
Moreover, the origin of the result of Ref.~\cite{Shlomo03} is unclear.
Does it correspond to an effective force with a symmetry energy
which is as stiff as that associated with the RMF Lagrangians ? Or
is some other parameter playing a role ?

It seems necessary to make a systematic analysis of what is the
upper limit for $K_\infty$ within the non-relativistic framework.
The value of 230 MeV is extracted from a subset of the existing
Skyrme parametrizations. Our main goal is to answer the question
whether it is possible to build new interactions in such a way
that $K_\infty$ becomes closer to the RMF value, as well as to
study what are the crucial quantities which control this variation
of $K_\infty$. In particular we would like to understand the (so
far, singular) result of Ref.~\cite{Shlomo03}.

The structure of the paper is the following. In Sec.~\ref{theo} we review how
the nuclear incompressibility is extracted from the microscopic calculations
using the ISGMR experimental data, and what are the plausible quantitative arguments
in favour of the idea that the density dependence of the symmetry energy plays a role.
In Sec.~\ref{skyrme} we describe our fitting of new Skyrme interactions suitable
for the microscopic ISGMR calculations, and in Sec.~\ref{results} we
describe the results obtained and the implications for the value of $K_\infty$.
In Sec.~\ref{conclu} we present our conclusions.

\section{ DEDUCING THE NUCLEAR INCOMPRESSIBILITY FROM MONOPOLE DATA AND
THE ROLE OF THE SYMMETRY ENERGY}
\label{theo}

In all the discussions of the relationship between the nuclear matter incompressibility
and the ISGMR in finite nuclei, the starting point is the definition given by
J.P. Blaizot\ \cite{Blaizot80} of the finite nucleus incompressibility $K_{\rm A}$ as
\begin{equation}
K_{\rm A} = {m \langle r^2 \rangle_0 E_{\rm ISGMR}^2\over \hbar^2},
\label{defKA}
\end{equation}
where $m$ is the nucleon mass and $\langle r^2 \rangle_0$ is the
ground-state mean square radius. This expression has a
well-defined meaning in medium-heavy nuclei, where the ISGMR is
associated to a single peak at the energy $E_{\rm ISGMR}\approx$
80${\rm A}^{-1/3}$. In light nuclei the monopole strength is very
much fragmented and many states show up, whose microscopic
structure does not correspond to the simple picture of the radial
``breathing mode'' according to theoretical 
calculations (see, for example, Ref.\ \cite{Gleissl}). In the
case of the nuclei studied in\ \cite{Youngblood99}, the existence
of a single, collective monopole state is quite evident from the
measured cross sections. In particular, in the case of $^{208}$Pb
which is the object of our present study, the experimental peak
energy and the centroid energies $E_0$ and $E_{-1}$ (defined
respectively as ${m_1\over m_0}$ and $\sqrt{m_1\over m_{-1}}$,
where $m_k$ is the $k$-th moment of the strength function)
essentially coincide, leaving out any ambiguity about the correct
value of $E_{\rm ISGMR}$ to be used for determining the
experimental value of $K_{\rm A}$.

However, finding a theoretical relation between $K_{\rm A}$ and $K_\infty$
is less simple. In Ref.\ \cite{Blaizot80}, the generic expression
of the energy functional associated with Skyrme-HF has been
written in the case of a finite spherical system. At variance with
that of infinite matter, the density is not uniform and cannot be
reduced to a simple number. Therefore, to minimize the energy
functional and find its second derivative around the minimum one
has to resort to various simplifying hypotheses. The main one is
the use of the so-called scaling model, in which a simple shape of
the ground-state density $\varrho_0$ is assumed and its changes
are associated to a single parameter $\lambda$, i.e., they are of
the type $\varrho_0(\vec r)\rightarrow\varrho_\lambda(\vec r)
={1\over\lambda^3}\varrho_0({\vec r\over\lambda})$. In this way,
the expression for the finite system incompressibility can be
found. By isolating the terms corresponding to the volume,
surface, symmetry and Coulomb contributions, the result can be
written as
\begin{equation}
K_{\rm A} = K_\infty + K_{surf}{\rm A}^{-1/3} + K_{sym}\delta^2 +
K_{Coul}{\rm Z^2\over \rm A^{4/3}},
\label{KAexpl}
\end{equation}
where $\delta\equiv (N-Z)/A$ (cf. Sec. 6.2 of Ref.\ \cite{Blaizot80}).

We remind here that, in the past many authors have used the
formula (\ref{KAexpl}) as an ansatz and have tried to obtain the
parameters of the r.h.s. from a numerical fit, using as input the
experimental values of $K_{\rm A}$ in different nuclei. This
procedure is not stable and leads to ill-defined values of the
parameters\ \cite{Pearson-SY}, so that it is nowadays abandoned.

Instead, the microscopic method to deduce $K_\infty$ relies on the fact that RPA
calculations of the ISGMR can be performed by using functionals characterized by
different values of $K_\infty$. If the calculations done with a given functional
reproduce the experimental ISGMR energy, the associated value of $K_\infty$ should
be chosen as the best one. Let us examine this in more detail.

Mainly one nucleus has been used so far, that is, $^{208}$Pb. In the first work
in which the microscopic procedure has been applied\ \cite{Blaizot95}, the RPA
values for $K_{\rm A}$ obtained from the RPA centroid energies $E_{-1}$ have been
plotted versus the $K_\infty$ of the force used. Then, an empirical 
linear fit of the results was performed, namely
\begin{equation}
K_{\rm A} = a K_\infty + b.
\label{linear}
\end{equation}
This relation allows to extract the best value for
$K_\infty$ by inserting the experimental $K_{\rm A}$. In\
\cite{Blaizot95} the explicit form of (\ref{linear}) in the case
of $^{208}$Pb is $K_{\rm A}=0.64 K_\infty - 3.5$ [MeV]. The second
term of the r.h.s. is much smaller than the first term.
Consequently, even if in principle the last formula together with
(\ref{defKA}) would lead to $E_{\rm ISGMR}=1.16\sqrt{0.64 K_\infty
- 3.5}$, this equation can be approximated by 
$0.928\sqrt{K_\infty}$ (neglecting the second term under the square root). 
This explains why in many of the works quoted in\ \cite{Colo03} a
successful interpolation of the type
\begin{equation}
E_{\rm ISGMR} = a' \sqrt{K_\infty} + b'
\label{linear2}
\end{equation}
was done: in practice, Eqs. (\ref{linear}) and (\ref{linear2}) are
equivalent. It is from either of these relations, using the
experimental ISGMR energy in $^{208}$Pb which is 14.17 $\pm$ 0.28
MeV, that the values for $K_\infty$ mentioned in Sec. I were
obtained.

The uncertainity of the value of $K_\infty$ which is deduced is
\begin{displaymath}
{\delta K_\infty \over K_\infty} = 2 {\delta E_{\rm ISGMR} \over E_{\rm ISGMR}}.
\end{displaymath}
The experimental error on the monopole energy, plus a theoretical error of the
same order (see\ \cite{Colo03}), produce a global error bar of $\pm$ 12 MeV on
$K_\infty$.

One may argue why the linear relations just introduced are valid.
So far, Eq. (\ref{KAexpl}) has not played in fact any explicit
role in the deduction of $K_\infty$. However, this expression can
be taken as a rather useful guideline. Given a microscopic
functional, the different terms $K_{surf}$, $K_{sym}$ and
$K_{Coul}$ (in addition to $K_\infty$) entering this formula
can be calculated as
described shortly below. The resulting value of $K_{\rm A}$
differs from the microscopic outcome of RPA, as a rule, by about
5\%. Therefore, we make in the rest of this Section a detailed
analysis of the role of the different terms in (\ref{KAexpl}). If,
for a family of functionals, $K_{\rm A}$ depends linearly on
$K_\infty$ as written in (\ref{linear}), it means that the other
terms do not vary significantly. This is what happens for a large
subset of the Skyrme and Gogny parametrizations, as it is evident
from Fig. 6 of Ref.\ \cite{Blaizot95}. However, the role of the
surface, symmetry and Coulomb terms should be critically
re-examined if new functionals, including relativistic ones, enter
in the discussion.

The expression for these terms have been given in Ref.\
\cite{Blaizot80}. $K_{surf}$ cannot be calculated analytically,
but numerical estimates are possible within both the quantal and
semiclassical scheme. We refer the reader to Ref.\
\cite{Pearson82} for an example of a quantal derivation (which is
a scaled Hartree-Fock calculation of semi-infinite nuclear
matter). The most recent semiclassical, i.e., extended
Thomas-Fermi (ETF) calculations, have been performed both in the
non-relativistic and in the relativistic scheme and have shown
that the quantity $K_{surf}$ is well approximated by $c K_\infty$
with $c\approx$ -1 (however, it should be noted that $c$ tends to
grow with $K_\infty$)\ \cite{Patra02}. We have checked that this
approximation is valid in the case of all the forces used in this
work: we have seen that, e.g., $c=-1.03$ if $K_\infty$=230 MeV and
$c=-1.07$ if $K_\infty$=250 MeV.

In order to study $K_{sym}$, we first give some necessary definitions of the symmetry
energy and of the parameters related to its density dependence. We define the symmetry
energy by writing the total energy density $\cal E$ as the sum of an isoscalar part
${\cal E}_0(\varrho)$ which depends only on the total density
$\varrho\equiv\varrho_n+\varrho_p$, and an isovector part,
\begin{equation}
{\cal E}(\varrho,\varrho_-) = {\cal E}_0(\varrho) + \varrho S(\varrho)
({\varrho_-\over\varrho})^2,
\end{equation}
where $\varrho_-\equiv\varrho_n-\varrho_p$. We remind in this context that in a
homogeneous system, $E/A={\cal E}/\varrho$. The symmetry energy $S(\varrho)$ can be
expanded up to second order around $\varrho_0$,
\begin{equation}
S(\varrho) = S(\varrho_0) + S'(\varrho_0) (\varrho-\varrho_0)
+ {1\over 2}S''(\varrho_0) (\varrho-\varrho_0)^2.
\end{equation}
The value of the symmetry energy at saturation $S(\varrho_0)$ is often denoted as
$J$ and we
are following
use the same notation in this paper. Other notations, like
$a_\tau$ or $a_4$, are also employed in the literature. The first
and second derivatives of $S(\varrho)$ at the saturation point
have been written many times in terms of the so-called parameters
$L$ and $K_{sym}$ (see, e.g., Ref.\ \cite{Treiner86}), as
$S'(\varrho_0)= L/3\varrho_0$ and
$S''(\varrho_0)=K_{sym}/9\varrho_0^2$. It is quite unfortunate
that the symbol $K_{sym}$ has been used in the literature with
such different meanings, either in connection with
$S''(\varrho_0)$ or in Eq. (\ref{KAexpl}). Here, we will always
use $K_{sym}$ to mean the symmetry term of $K_{\rm A}$ in Eq.(3).

The expression of $K_{sym}$ is
\begin{equation}
K_{sym} = 9 \varrho_0^2 S''(\varrho_0) + 9 \varrho_0 S'(\varrho_0) -
{81 \varrho_0^3 S'(\varrho_0) \over K_\infty} {d^3{\cal E}\over d\varrho^3}\vert_{\varrho_0},
\label{Ksym}
\end{equation}
and from this expression it is evident that this parameter contains some relevant
information about the density dependence of the symmetry energy.

The values of $J$ are, as a rule, larger in the case of the RMF
functionals than for the Skyrme ones. A larger value of $J$ is
correlated with a larger value of $S'(\varrho_0)$, which is
usually a positive quantity although it may sometimes become
negative (cf.\ \cite{Oyamatsu02} and Fig. 4 of\ \cite{Vretenar03},
as well as Tables\ \ref{table:properties1} and\
\ref{table:properties2} in this paper). In the next Section we
show that a larger $J$ is also correlated with a more negative
value of $K_{sym}$, at least for the forces we have studied. The
explanation which is given for the correlation between $J$
and $S'(\varrho_0)$ is that the fits to finite nuclei observables
constrain the symmetry energy at some average density $\langle
\varrho \rangle$ lower than $\varrho_0$ (see, e.g., Ref.\
\cite{Furnstahl02} and references therein). In the case of one set of
forces introduced in this paper (see Sec.\ \ref{skyrme}), this
typical behaviour of the symmetry energy is shown in Fig.\
\ref{Figure_sym}. In a narrow region around $\varrho$=0.10
fm$^{-3}$ ($\pm$ 0.001 fm$^{-3}$) all curves cross one another at
a value $S(\varrho)$=25 $\pm$ 1 MeV.
When the symmetry energy at saturation is larger, the slope is
also larger. The other sets of forces show qualitatively the same
trend.

The last term of (\ref{KAexpl}) is the Coulomb contribution which is unlikely
to be very much model dependent. It is written as
\begin{equation}
K_{Coul} = {3\over 5}{e^2\over r_0} (1-{27\varrho_0^2\over K_\infty}
{d^3{\cal E}\over d\varrho^3}\vert_{\varrho_0}),
\label{KCoul}
\end{equation}
where $r_0$ is the average inter-particle spacing.

In summary,
if we want to compare two models, say I and II (they could be for
instance a non-relativistic and a RMF functional, respectively),
we will write, by using $K_{surf}=cK_\infty$,
\begin{eqnarray}
K_{\rm A} & \sim &
K_\infty^{({\rm I})}  (1+c{\rm A}^{-1/3}) + K_{sym}^{({\rm I})}\delta^2
+ K_{Coul}^{({\rm I})} {{\rm Z}^2\over {\rm A}^{4/3}},
\nonumber \\
K_{\rm A} & \sim &
K_\infty^{({\rm II})} (1+c{\rm A}^{-1/3}) + K_{sym}^{({\rm II})}\delta^2
+ K_{Coul}^{({\rm II})} {{\rm Z}^2\over {\rm A}^{4/3}}.
\label{KAcompare}
\end{eqnarray}
We have already mentioned that $K_{sym}$ is negative, and the same
is true for $K_{surf}$ and $K_{Coul}$. All can be viewed as
corrections to the leading term $K_\infty$. It is clear that {\it
a more negative value of $K_{surf}$ or $K_{sym}$ leads to
extracting from the experimental $K_{\rm A}$ a larger value of
$K_\infty$.} We will develop this argument in Sec.\ \ref{results}.

\section{CONSTRUCTION OF NEW SKYRME PARAMETER SETS}
\label{skyrme}

The different forces used in this study have been built using a procedure which is quite
similar to the one discussed in Ref.~\cite{Chabanat98}. The starting point is the standard
form of a Skyrme interaction as given in Eq. (2.1) of~\cite{Chabanat98}.

In the case of the first set of forces that we have constructed,
the density-dependent term has $\varrho^{\alpha}=\varrho^{1/6}$.
The spin-gradient terms occuring in the Skyrme functional are
neglected and the Coulomb exchange term is included within the
Slater approximation. The center-of-mass motion is taken into
account with the usual ${{\rm A}\over {\rm A - 1}}$ correction in
the kinetic term,
which means that only the one-body part of the center-of-mass
energy is subtracted before variation.

The parameters of the forces have been determined by minimizing a $\chi^2$ built on:
\begin{enumerate}
\item the infinite nuclear matter properties $\varrho_0$,
$E/A(\varrho_0)$ (while $K_{\infty}$, $J$ and the enhancement
factor $\kappa$ of the Thomas-Reiche-Kuhn sum rule are kept
constant); \item the following finite nuclei properties: binding
energies and charge radii of $^{40,48}$Ca, $^{56}$Ni and
$^{208}$Pb together with the binding energy of $^{132}$Sn; \item the
spin-orbit splitting of the neutron $3p$ shell in $^{208}$Pb;
\item the surface energy, calculated in the ETF approximation and
fitted to the value of the SkM$^*$ force in order to obtain good
mean field properties at large deformations (especially a good fission
barrier of the $^{240}$Pu nucleus).  
\end{enumerate}
Furthermore, the parameter $x_2$ is fixed to $-1$ in order to
ensure the stability of the fully polarized neutron matter in a
simple but tractable way\ \cite{Kutschera94}.
Unlike the case of the SLy4 force, the equation of state of
neutron matter is checked but not fitted in order to have a large
enough variational space of parameters when the nuclear
incompressibility and the symmetry energy are varied. The forces
which have been built have $K_\infty$ equal to 230, 240 and 250
MeV whereas $J$ is varied between 26 and 40 MeV. In Fig.\
\ref{Figure_acc1} we show the accuracy of the present forces in
reproducing the ground-state observables (binding energies and
charge radii).

Motivated by the comparison with Ref.\ \cite{Shlomo03}, we have
also built another set of forces with a similar protocol but with
the density dependent term $\varrho^{\alpha}$ having the same 
exponent $\alpha$=0.3563 as the force SK255 introduced in 
Ref.\ \cite{Shlomo03}. 
The forces of this set have $K_\infty$ equal to 250, 260 and 270 MeV
while $J$ is varied between 28 MeV and 42 MeV. Fig.\
\ref{Figure_acc2} gives an idea of the accuracy of this set, in
the same way as for the previous one.

The nuclear matter properties associated to all the new forces introduced
in this paper are summarized in Tables\ \ref{table:properties1}
and\ \ref{table:properties2}. By looking at the values of the effective 
mass, one can recognize the well-known correlation between
$K_\infty$, $\alpha$ and $m^*/m$ (see, e.g., Fig. 2 of 
Ref.\ \cite{Chabanat98}).    

\section{RESULTS AND DISCUSSION}
\label{results}

Using these new Skyrme interactions, the ISGMR centroid energies
$E_{-1}=\sqrt{m_1\over m_{-1}}$ in $^{208}$Pb have been calculated
in a fully self-consistent manner. The energy-weighted sum rule
$m_1$ is obtained from the well-known double commutator
expectation value, while the inverse energy-weighted sum rule
$m_{-1}$ is extracted by means of a constrained HF (CHF)
calculation\ \cite{Bohigas79}. Adding to the Hamiltonian a term
$\lambda\hat M$, where $\hat M$ is in this case the monopole
operator $\sum_{i=1}^A r_i^2$ and $\lambda>0$ (to avoid an
Hamiltonian without lowest bound), the value of $m_{-1}$ can be
extracted in two different ways, that is,
\begin{equation}
m_{-1} = -{1\over 2}{\delta \langle M \rangle_0 \over \delta \lambda} =
{1\over 2}{\delta^2 \langle H \rangle_0 \over \delta \lambda^2}.
\end{equation}
By varying the steps in
$\lambda$ and by comparing the outcome of these two different
expressions, numerical tests concerning the accuracy of
$m_{-1}$ can be performed. We have
come to the conclusion that this quantity can be determined with
an accuracy of $\pm$ 3\% or better. This is definitely more
reliable than the result of usual RPA calculations made using a
basis expansion, since the convergence of $m_{-1}$ with the basis
size can be quite slow. Moreover, as already discussed in Sec.\
\ref{intro}, Skyrme RPA calculations of the ISGMR performed so
far lack full self-consistency since part of the residual
interaction (the two-body Coulomb and two-body spin-orbit terms)
are dropped. This has been shown to lead to a systematic error in
the monopole centroid energies\ \cite{Colo03}.

In Figs.\ \ref{Figure_res1a} and\ \ref{Figure_res1b} we show the
results for the monopole energy $E_{-1}$ obtained with the present
interactions, as a function of the associated values of $K_\infty$
and $J$.
Fig.\ \ref{Figure_res1a} refers to the forces with $\alpha$=1/6,
whereas Fig.\ \ref{Figure_res1b}
is for those with $\alpha$=0.3563. 
In the case of the forces with $\alpha$=0.3563, we have used the same
procedure of Ref.\ \cite{Shlomo03}, that is, we have 
neglected the Coulomb exchange and we have omitted the 
center-of-mass correction in the HF
variation, by 
subtracting it afterwards from the total energy, in the harmonic
oscillator approximation ($E_{c.m.} = {3\over 4}$41A$^{-1/3}$).
We have checked that this lowers the monopole energy
by about 150 keV. 
The straight lines in the
figures are linear fits of the CHF results corresponding to the
different symbols, whereas the experimental range for the monopole
energy\ \cite{Youngblood99}
is delimited by the horizontal lines. These figures can be
compared with Fig. 5 (upper panel) of Ref.\ \cite{Vretenar03}. The
results for the monopole energy are, as expected, much less
sensitive to $J$ than to $K_\infty$. By varying $K_\infty$ by 10
MeV, i.e., by about 4\%, the monopole energy changes by 0.5 MeV.
In order to obtain the same change, $J$ should be varied from 26
MeV to 40 MeV which is about 50\%. The RMF results show
qualitatively the same pattern.

We have to stress that the existence of a definite, yet not
strong, dependence on $J$ is in agreement with the discussion in
Sec.\ \ref{theo}, where the role of $K_{sym}$ as one of the
crucial parameters governing the monopole energy has been
emphasized. It is clear from Figs.\ \ref{Figure_res2a} and\
\ref{Figure_res2b} that the monopole energies do depend on the
parameter $K_{sym}$, associated with the density dependence of the
symmetry energy. On the other hand, in the forces we have built,
there is a strong correlation between $K_{sym}$ and $J$,
essentially independent of $K_\infty$ but not of $\alpha$. This
correlation is shown in Fig.\ \ref{Figure_corr}. The
interpretation of this plot is the following. The modification of
the exponent $\alpha$ in the Skyrme functional, allows us to
change the value of $K_{sym}$ keeping fixed the values of
$K_\infty$ and $J$. According to the argument developed at the end
of Sec.~\ref{theo}, this should allow to change the value of
$K_\infty$ extracted from the experimental ISGMR data.

By considering only the set with $\alpha$=1/6, we confirm the
previous result of Ref.\ \cite{Colo03} that $K_\infty$=230-240 MeV is
the preferred value for the nuclear incompressibility. This is not
fully compatible with the RMF result. In fact, extrapolating from
Fig.\ \ref{Figure_res1a}, one can see that an hypothetic Skyrme
parametrization having that associated value of $K_\infty$, would
reproduce the experimental monopole energy only with an
unrealistic value of $J$ above 50 MeV. On the other hand, the set
of forces with $\alpha$=0.3563 allows extracting a value of
$K_\infty$ around 250 MeV, in agreement with the outcome of the
RMF calculations.

We should at this point stress that the more negative values of
$K_{sym}$ which characterize the forces with $\alpha$=0.3563
cannot, alone, explain the extraction of a larger $K_\infty$. The
forces which better reproduce the experimental monopole energy are
those with ($\alpha$, $K_\infty$, $J$) given either by (1/6, 230
MeV, 28 MeV) or by (0.3563, 250 MeV, 30 MeV). We can apply Eq.
(\ref{KAcompare}) to the case of these two forces, by inserting
the values given in the Tables\ \ref{table:properties1} and\
\ref{table:properties2} and by taking into account that $c\sim$ -1 
for $K_\infty$=230 MeV but $\sim$ -1.1 for
$K_\infty$=250 MeV. This gives for the two forces, respectively,
\begin{displaymath}
K_{\rm A} = 154.16 = 230 - 38.82 - 10.23 - 26.79\ \rm [MeV]
\end{displaymath}
and
\begin{displaymath}
K_{\rm A} = 160.64 = 250 - 46.41 - 15.18 - 27.77\ \rm [MeV],
\end{displaymath}
where the four numbers on the r.h.s. correspond respectively to
the volume, surface, symmetry and Coulomb contributions. It is
clear that most of the gain of 20 MeV in $K_\infty$ comes from the
increase of the surface term (8 MeV), and the more negative
$K_{sym}$ which, multiplied by the tiny value of $\delta^2$ for
$^{208}$Pb, contributes by 5 MeV. A small contribution of 1 MeV
results from the increase of $K_{Coul}$. Finally, Eq.
(\ref{KAcompare}) does not consider that, in the calculations done
by employing the forces with $\alpha$=0.3563 the Coulomb exchange
and center-of-mass corrections are neglected. As mentioned above,
this lowers the ISGMR energies by about 150 keV and
hence $K_{\rm A}$ by about 5 MeV. This brings the two results for
$K_{\rm A}$ rather close to each other.

\section{CONCLUSION}
\label{conclu}

Until recently, the extraction of the nuclear incompressibility from the monopole
data was plagued by a marked model dependence: the Skyrme energy functionals seemed
to point to 210-220 MeV, the Gogny functionals to 235 MeV, and the relativistic
functionals to 250-270 MeV.
It has been shown in Ref.\ \cite{Colo03} that the result
of the Skyrme functionals is in fact consistent with that of
Gogny, i.e., 235 MeV using the $^{208}$Pb data. The previous value
of 210-220 MeV was derived using non fully self-consistent
calculations neglecting the residual Coulomb and spin-orbit
interactions. The discrepancy between the non-relativistic value
of 235 MeV and the relativistic prediction remained, since new
relativistic calculations confirmed the lower bound of about 250
MeV.

The work of Agrawal {\it et al.}\ \cite{Shlomo03} suggests that it
is possible to build Skyrme forces which fit nuclear ground states
and lead to the correct monopole energy, with $K_\infty$=255 MeV.
Very recently, Goriely {\it et al.}\ \cite{Goriely} have also obtained
Skyrme parametrizations which fit very well the binding energies
of a quite large number of nuclei. Some of the parametrizations
having ($\alpha$ =0.5, $m^*/m$ =0.80, $J$ =30 MeV, $K_\infty$ =276 MeV)
are doing satisfactorily well for the ISGMR energy in $^{208}$Pb
($E_{\rm ISGMR} \simeq$ 14 MeV). In the present work we systematically
explore what are the conditions which lead to such different
results for $K_\infty$ within the Skyrme framework.

To this aim, we build classes of Skyrme forces which span a wide range of values
for $K_\infty$ and for the symmetry energy at saturation $J$. All these forces
reproduce the ground state observables with good accuracy.
We use them to calculate the monopole energy in $^{208}$Pb, defined as
$E_{-1}=\sqrt{m_1\over m_{-1}}$. We stress again that we can obtain this quantity without any lack
of self-consistency, and with a numerical error which is not larger
than the experimental
uncertainity.

A first class of forces are built using the SLy4 protocol and have
a density dependence characterized by the exponent $\alpha$=1/6.
With these forces, a value of $K_\infty$ around 230-240 MeV is
obtained, confirming the previous results of\ \cite{Colo03}. To
obtain the correct monopole energy with larger values of
$K_\infty$ would require an unrealistically large value of $J$,
since $E_{\rm ISGMR}$ increases with $K_\infty$ and decreases with
$J$. We understand this latter dependence as a consequence of the
direct relation between the $K_\infty$ and the {\em density
dependence} of the symmetry energy $S(\varrho)$, and of the
unavoidable correlation between $S(\varrho)$ and $J$.

To solve the discrepancy with the result of Agrawal {\it et al.},
we have built a second class of forces which have the
density dependence $\alpha$=0.3563. Using this class of forces 
we can arrive at $K_\infty$ between 250 and 260 MeV.
Actually, we can reproduce very accurately the results of Ref.\
\cite{Shlomo03}, if we use the same approximations made in that
work, namely if we neglect the Coulomb exchange
and center-of-mass corrections in the HF mean field. This shows
that the differences between Ref.\ \cite{Shlomo03} and our work 
in the detailed protocol used to determine the forces, are 
unimportant. We have observed that the differences between 
the results of the two classes of Skyrme forces built in the 
present paper, come both from the surface and
symmetry contributions, as a consequence of the change in the
exponent $\alpha$, and from the neglect of Coulomb exchange and
center-of-mass corrections, which affect the monopole energy by
about 150 keV and therefore, $K_\infty$ by about 5 MeV.

In conclusion, within the non-relativistic framework there is not a 
unique relation between the value of $K_\infty$ associated with an 
effective force and the monopole energy predicted by that force. 
{\it Bona fide} Skyrme forces can either predict 230-240 MeV for 
$K_\infty$ or arrive at 250 MeV if a different density
dependence is adopted and if one excludes some terms from the energy
functional. This latter procedure, although it may mimick the relativistic case,
is not conceptually well justified.

\section*{ACKNOWLEDGMENTS}

One of us (G.C.) has the pleasure of acknowledging deep and enlightening discussions with
M. Centelles and X. Vi\~{n}as about the CHF calculations.

\newpage

\begin{table}
\caption{Nuclear matter properties calculated with
the different Skyrme parameter sets characterized by $\alpha$=1/6 and
by different values of $K_\infty$ and $J$ (these two quantities identify the
parameter set and are shown in the first column). All quantities are defined
in the text.}
\label{table:properties1}
\begin{ruledtabular}
\begin{tabular}{ccccccc}
& $\varrho_0$ & $E/A$ & $m^*/m$ & $L$
& $K_{sym}$   & $ K_{Coul}$ \\
& [fm$^{-3}$] & [MeV] &         & [MeV]
& [MeV]       & [MeV] \\
\hline
230/26     & 0.161 & -15.89 & 0.69   & -39.06
& -178.95     & -4.92      \\
230/28     & 0.162 & -15.96 & 0.70   & -11.23
& -228.62     & -4.91      \\
230/30     & 0.161 & -15.98 & 0.70   &  22.88
& -281.05     & -4.90      \\
230/32     & 0.161 & -16.03 & 0.70   &  36.22
& -314.95     & -4.90      \\
230/34     & 0.161 & -16.06 & 0.70   &  56.15
& -354.24     & -4.89      \\
230/36     & 0.161 & -16.10 & 0.71   &  71.55
& -389.53     & -4.88      \\
230/38     & 0.161 & -16.14 & 0.71   &  87.60
& -424.86     & -4.88      \\
230/40     & 0.161 & -16.16 & 0.71   & 106.07
& -462.20     & -4.87      \\
\hline
240/26     & 0.160 & -15.91 & 0.63   & -15.95
& -176.41     & -5.05      \\
240/28     & 0.161 & -15.94 & 0.63   &  3.97
& -202.07     & -5.04      \\
240/30     & 0.159 & -15.95 & 0.63   &  34.05
& -273.93     & -5.04      \\
240/32     & 0.166 & -16.15 & 0.65   &  34.43
& -300.39     & -5.00      \\
240/34     & 0.164 & -16.12 & 0.65   &  62.60
& -350.02     & -5.01      \\
240/36     & 0.164 & -16.15 & 0.65   &  75.67
& -384.99     & -5.00      \\
240/38     & 0.163 & -16.19 & 0.65   &  98.62
& -429.77     & -5.00      \\
240/40     & 0.165 & -16.24 & 0.65   & 108.15
& -460.88     & -4.99      \\
\hline
250/28     & 0.165 & -16.10 & 0.59   &  32.99
& -238.76     & -5.13      \\
250/30     & 0.164 & -16.09 & 0.59   &  30.02
& -255.78     & -5.13      \\
250/32     & 0.164 & -16.14 & 0.59   &  43.59
& -293.94     & -5.12      \\
250/34     & 0.163 & -16.14 & 0.59   &  60.33
& -334.94     & -5.12      \\
250/36     & 0.162 & -16.17 & 0.59   &  80.19
& -379.80     & -5.12      \\
250/38     & 0.162 & -16.20 & 0.59   &  97.50
& -421.63     & -5.12      \\
250/40     & 0.162 & -16.25 & 0.59   & 112.18
& -460.37     & -5.11      \\
\end{tabular}
\end{ruledtabular}
\end{table}

\newpage

\begin{table}
\caption{The same as Table\ \ref{table:properties1} for
the different Skyrme parameter sets characterized by $\alpha$=0.3563.}
\label{table:properties2}
\begin{ruledtabular}
\begin{tabular}{ccccccc}
& $\varrho_0$ & $E/A$ & $m^*/m$ & $L$
& $K_{sym}$   & $ K_{Coul}$ \\
& [fm$^{-3}$] & [MeV] &         & [MeV]
& [MeV]       & [MeV] \\
\hline
250/30     & 0.153 & -15.87 & 0.77   &  12.66
& -339.16     & -5.09      \\
250/32     & 0.152 & -15.89 & 0.77   &  36.57
& -377.81     & -5.09      \\
250/34     & 0.151 & -15.91 & 0.77   &  58.81
& -415.36     & -5.08      \\
250/36     & 0.150 & -15.92 & 0.77   &  72.00
& -447.51     & -5.09      \\
250/38     & 0.149 & -15.95 & 0.77   &  95.20
& -485.39     & -5.08      \\
250/40     & 0.148 & -15.96 & 0.77   & 110.19
& -518.58     & -5.08      \\
250/42     & 0.148 & -15.97 & 0.77   & 126.53
& -552.32     & -5.08      \\
\hline
260/28     & 0.157 & -15.96 & 0.67   &  -7.30
& -318.87     & -5.20      \\
260/30     & 0.157 & -16.00 & 0.68   &  16.92
& -326.92     & -5.19      \\
260/32     & 0.159 & -16.08 & 0.69   &  29.53
& -362.73     & -5.17      \\
260/34     & 0.159 & -16.13 & 0.70   &  46.05
& -399.70     & -5.16      \\
260/36     & 0.160 & -16.19 & 0.71   &  64.11
& -437.67     & -5.15      \\
260/38     & 0.160 & -16.25 & 0.72   &  78.40
& -471.97     & -5.14      \\
260/40     & 0.161 & -16.30 & 0.73   &  90.97
& -505.06     & -5.13      \\
\hline
270/28     & 0.157 & -15.97 & 0.58   &  -4.60
& -262.94     & -5.32      \\
270/30     & 0.157 & -16.01 & 0.58   &  21.47
& -311.34     & -5.31      \\
270/32     & 0.157 & -16.05 & 0.58   &  43.36
& -355.80     & -5.30      \\
270/34     & 0.157 & -16.10 & 0.59   &  63.76
& -398.68     & -5.29      \\
270/36     & 0.157 & -16.14 & 0.59   &  81.41
& -438.67     & -5.29      \\
270/38     & 0.158 & -16.19 & 0.60   &  98.06
& -477.62     & -5.28      \\
270/40     & 0.157 & -16.21 & 0.60   & 115.06
& -516.21     & -5.27      \\
270/42     & 0.157 & -16.23 & 0.60   & 133.90
& -556.22     & -5.27      \\
\end{tabular}
\end{ruledtabular}
\end{table}

\newpage

\begin{figure}
\caption{Density dependence of the symmetry energy 
for one of the set of forces
($\alpha$=1/6 and $K_\infty$=240 MeV). The symmetry energy 
$J$ is varied from 26 to 40 MeV.}
\label{Figure_sym}
\end{figure}

\begin{figure}
\caption{Difference
between experimental binding energies (left) or experimental
charge radii (right) with the predictions of forces characterized
by $\alpha$=1/6 and by different values of $K_\infty$ and $J$,
for typical spherical nuclei. Note that the binding energies of
$^{40,48}$Ca, $^{56}$Ni, $^{132}$Sn and $^{208}$Pb, as well as the
charge radii of $^{40,48}$Ca, $^{56}$Ni and $^{208}$Pb are used in
the fit of the force parameters.}
\label{Figure_acc1}
\end{figure}

\begin{figure}
\caption{Same as Fig.\ \ref{Figure_acc1} in the case of the forces
with $\alpha$=0.3563.}
\label{Figure_acc2}
\end{figure}

\begin{figure}
\caption{The $^{208}$Pb ISGMR centroid energy $E_{-1}$ calculated
with the Skyrme parameter sets with $\alpha$=1/6, 
as a function of $J$. The different
symbols correspond to the values of $K_\infty$ (see inset). 
Lines are numerical fits and are simply intended to guide the
eye. The area delimited by the two horizontal lines correspond to the
experimental value.}
\label{Figure_res1a}
\end{figure}

\begin{figure}
\caption{Same as Fig.\ \ref{Figure_res1a}, for the
Skyrme parameter sets with $\alpha$=0.3563.}
\label{Figure_res1b}
\end{figure}

\begin{figure}
\caption{Same as Fig.\ \ref{Figure_res1a}, with the results
plotted as a function of $K_{sym}$.}
\label{Figure_res2a}
\end{figure}

\begin{figure}
\caption{Same as Fig.\ \ref{Figure_res1b}, with the results
plotted as a function of $K_{sym}$.}
\label{Figure_res2b}
\end{figure}

\begin{figure}
\caption{Correlation plot $K_{sym}$ vs. $J$. Squares and triangles display
respectively the results for the forces with $\alpha$=1/6 and 0.3563. 
Straight lines are numerical fits.}
\label{Figure_corr}
\end{figure}

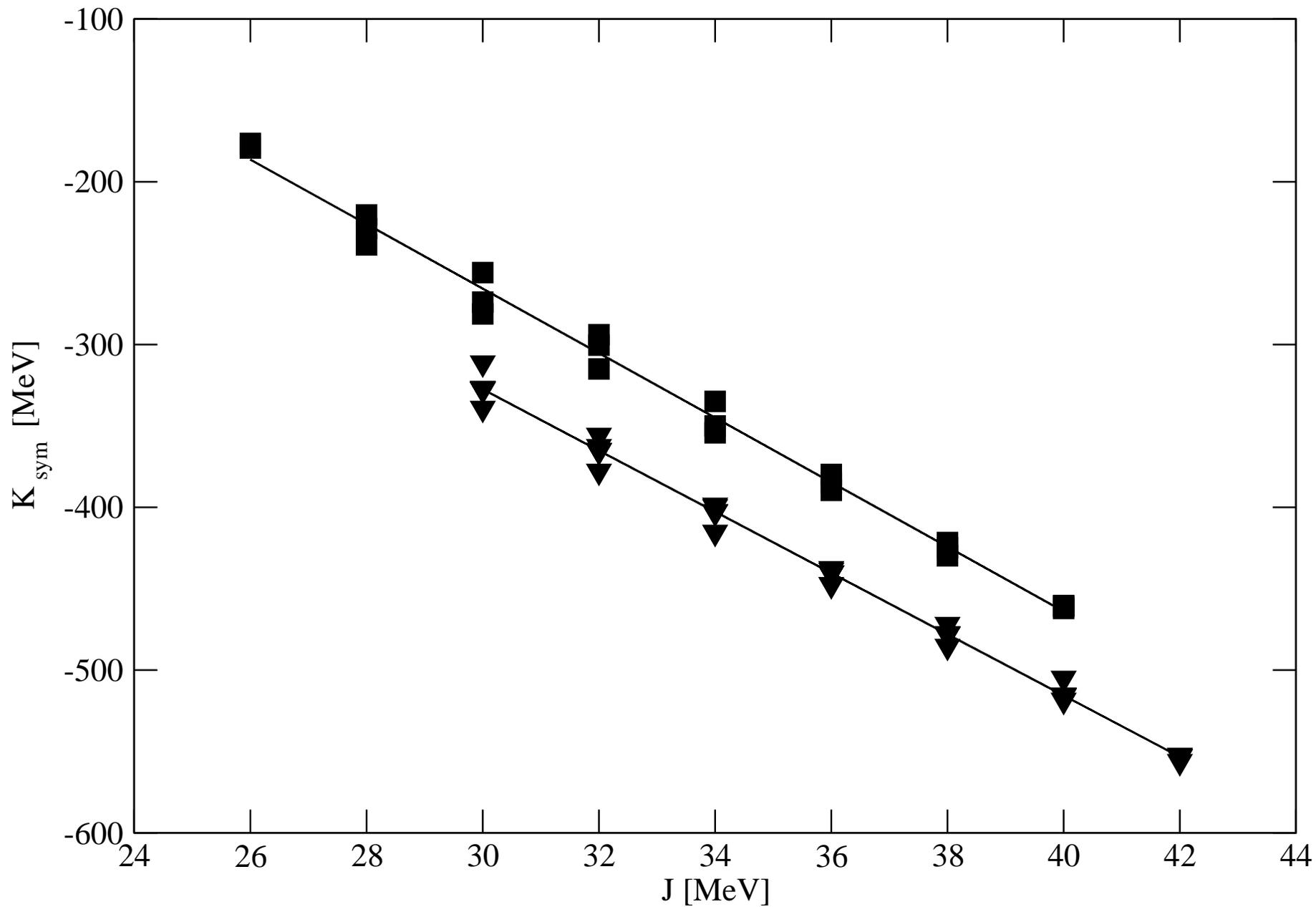
\end{document}